\documentclass[12pt]{article}
\usepackage{graphicx}
\newcommand{\bea}{\begin{eqnarray}}
\newcommand{\eea}{\end{eqnarray}}
\newcommand{\be}{\begin{equation}}
\newcommand{\ee}{\end{equation}}
\newcommand{\vs}[1]{\vspace{#1 mm}}

\newcommand{\dsl}{\pa \kern-0.5em /}

\newcommand{\pa}{\partial}

\newcommand{\nn}{\nonumber\\}

\newcommand{\umatrix}{\rlap 1\mkern4mu{\rm l}}
\begin{document}
\topmargin 0pt
\oddsidemargin 0mm

\begin{flushright}
\end{flushright}

\vs{0.5}
\begin{center}
{\Large \bf Generating a dynamical M2 brane   \\
\vspace{2mm}
 from super-gravitons
  in a pp-wave background} \vs{10}

{\large Yi-Fei Chen, J. X. Lu and Nan Zhang
 }

 \vspace{5mm}

{\em
  Interdisciplinary Center for Theoretical Study\\
 University of Science and Technology of China, Hefei, Anhui 230026,
P. R. China\\
and\\
 Peng Huanwu Center for Fundamental Theory, Hefei, Anhui 230026, China\\
}
\end{center}

\vs{5}
\centerline{{\bf{Abstract}}}
\vs{5}
\begin{small}
We present a detail study of dynamically generating a M2 brane
from super-gravitons (or D0 branes) in a pp-wave background
possessing maximal spacetime SUSY. We have three kinds of
dynamical solutions depending on the excess energy which appears
as an order parameter signalling a critical phenomenon about the
solutions. As the excess energy is below a critical value, we have
two branches of the solution, one can have its size zero while the
other cannot for each given excess energy. However there can be an
instanton tunnelling between the two. Once the excess energy is
above the critical value, we have a single solution whose
dynamical behavior is basically independent of the background
chosen and whose size can be zero at some instant. A by product of
this study is that the size of particles or extended objects 
can grow once there is a non-zero excess energy even without the
presence of a background flux, therefore lending support to the
spacetime uncertainty principle.
\end{small}
\newpage
\section{Introduction}
In our previous work\cite{clone}, we find a dynamical creation and
annihilation of a spherical D2 brane from N D0 branes via a
background flux which extends earlier work of the static creation
of a higher dimensional brane from lower dimensional ones via a
background flux \cite{myersone,dtvone}. In this case, the
background satisfies the bulk equations of motion only to leading
order when the $N$ is large. In the present work, we try to study
a similar problem but in a pp-wave background which is obtained
from the pp-wave limit of either $AdS_4 \times S^7$ or $AdS_7
\times S^4$, and satisfies the bulk equations of motion exactly.
In other words, we try to find how a spherical M2 brane can be
generated dynamically from N super-gravitons (or D0 branes) once
there is a non-zero excess energy. We still have three kinds of
dynamical solutions as in \cite{clone} but now because of the very
pp-wave background, two of the solutions differ from what we had
there. For example, we no longer have the analogue of a photon
creating an electron-positron pair in the presence of a background
and then annihilating back to a photon. We don't have the analogue
of a finite size dynamical configuration with negative excess
energy. We actually always need a positive excess energy to have a
dynamical configuration in the present case. We have here two branches
of solution for each given excess energy when it is below a
certain critical value but in \cite{clone} we
have only one solution for each given allowed excess energy. We
have there only one non-trivial finite size but preserving no supersymmetry (SUSY) static configuration while in present case we have two SUSY
preserving BPS static configurations, one is the zero-size
super-gravitons and the other  the finite size giant. In spite of
these differences, there are still many similarities. For example,
in both cases, the value of excess energy characterizes the
behavior of the underlying dynamics. SUSY can be preserved only
for vanishing excess energy. There is a transition behavior in
both cases that signifies either a finite size becoming zero or
the change of branches of dynamical solution. At the transition
value of the excess energy, the dynamical solution(s) can always
be expressed in terms of elementary functions (rather than using,
for example, Jacobian elliptic functions).

    In this paper, we have a dynamical spherical M2 brane whose
size cannot be zero when the positive excess energy is below the
critical value.  When the excess energy approaches zero, this
solution reduces to the SUSY preserving BPS giant which was
discovered in the matrix theory context in \cite{bmnone}. This
dynamical spherical M2 brane with a finite size is analogous to
the finite size dynamical spherical D2 brane with negative excess
energy given in \cite{clone} in the sense that both sizes cannot vanish
classically. However, for the present case, we have in addition
another dynamical solution with the same excess energy but its
size can be zero at some instant. In particular, when the excess
energy approaches zero, it reduces to zero-size SUSY preserving
BPS super-gravitons. This configuration is degenerate with the
giant one, therefore we expect instanton tunnelling between the
two even though classically the two are completely disconnected.
In other words, we still expect that the dynamical spherical M2
brane with a finite size can turn itself to the one with zero size
at some instant by instanton tunnelling, i.e., by quantum
mechanical effect, while the one discussed in \cite{clone} has no
such a luxury.

As pointed out in \cite{gmtone}, a spherical brane can be viewed
as a semi-spherical brane--anti semi-spherical brane pair. However
such a system differs from the usual infinitely extended
brane--anti brane pair in that SUSY can sometimes be preserved for
the former, for examples,  many of the giant configurations found
in \cite{mstone,gmtone,hhione} while the latter always breaks SUSY
and is unstable with the appearance of tachyon
modes\cite{asone,astwo}. The reason for such difference is that
for those cases when SUSY preserving BPS giants can be formed,
there are no excess energies involved  while the latter always
involve negative excess energies. In other words, such finitely
extended brane-anti brane system may serve the link between the general brane--anti brane systems and the associated 
tachyon condensations by Sen and others \cite{asone,astwo,yione}
and the SUSY preserving giant gravitons\cite{mstone,gmtone,hhione}
plus the other related ones\cite{dtvone,djmone}.

From the above, we see that the excess energy plays an important
role in determining whether a configuration is SUSY preserving or
not. This is not surprised at all since the excess energy measures
whether the corresponding BPS bound is saturated. Only for zero
excess energy, we can have the bound saturated which is necessary
for preserving SUSY. In other words, we have to find means to get
rid of the excess energy with respect to a relevant SUSY
preserving configuration so that the system under consideration
becomes SUSY preserving BPS one.

 It appears at present that in
order to generate a finite size, for example, from gravitons (or
D0 branes), we need either a relevant background flux (Myers
effect) or angular momentum plus a backgound flux. In other words, the generated finite size seems
necessarily related to a background flux. This appears also the case
for the appearance of non-commutativity in string
theory\cite{cdsone,dhone,swone,chone,bsone}. However, the
spacetime uncertainty principle proposed in
\cite{tyetalone,susskindone} seems a general one which has nothing
to do with the appearance of a background flux and it basically
says that when one increases one's time resolution, any probe
available will increase its physical size. Putting in another way,
a particle size will grow with its energy. We try also in this
paper to resolve the above apparent puzzle. As we will see that
the need to have a background flux for generating a finite size
from a zero one is merely an artifact and this is entirely due to
the constraint of finding either a static or a stationary (BPS)
configuration. We will show in section 2 that any time a particle
or an extended object has an excess energy above the corresponding
BPS bound, then its size will grow and no background flux is
needed. Therefore we provide a direct evidence for the spacetime
uncertainty principle\cite{tyetalone}.

    Other relevant or related work include those in \cite{total}.

     This paper is planned as follows: In section 2, we show
that by considering N super-gravitons described by the DLCQ matrix
theory, there will appear a finite size of spherical M2 brane if
there is a non-zero excess energy  in a flat background with no
flux, therefore lending support to the spacetime uncertainty
principle. In section 3, we will consider the DLCQ matrix theory
of N supergravitons in a pp-wave background and find that there
are three kinds of dynamical spherical M2 branes. We will discuss
the details and their differences from what we discussed in
\cite{clone}. Section 4 ends with a discussion of the implications
of the present work.

\section{Evidence for the spacetime uncertainty principle}

    For this purpose, we consider  the DLCQ matrix theory of N supergravitons with
the simplest possible flat background without any flux presence.
The bosonic part of the Lagrangian is \be\label{bsp} L = \frac{1}{2 R} {\rm Tr}\,
(\dot \Phi^i)^2 + \frac{R}{4} {\rm Tr}\, [\Phi^i, \Phi^j]^2, \ee where we
have chosen the eleven dimensional plank length $l_p = 1$, $R$
is the radius of the compactification along the light-like
direction $x^-$ and the dot above $\Phi^{i}$ denotes the time derivative. We also choose the gauge field $A_0 = 0$ in the
above. The equation of motion can be read from the above as \be
\ddot \Phi^i + R^2 [[\Phi^i, \Phi^j], \Phi^j] = 0,\ee which has
trivial static solution. We however look for a non-trivial
dynamical configuration which solves the above equation. Each of
the above $\Phi^i$ is a $N\times N$ matrix and it can easily be
examined that  $\Phi^i \sim \umatrix_{N\times N}$ is a trivial
solution. For the purpose of serving the need for finding evidence
in support of the spacetime uncertainty principle, we here choose
$\Phi^i$ for $i = 1, 2, 3$ each in the irreducible $N \times N$
representation of SU(2) with other $\Phi^j$ with $j = 4, \cdots
9$ as trivial. In other words, $\Phi^i = u(t) J^i$ with $J^i$ the
SU(2) generators in the $N \times N$ irreducible representation.
Therefore, we have from the above, using $[J^i, J^j] = {\rm i}\,
\epsilon^{ijk} J^k$, the following, \be \ddot u + 2 R^2 u^3 =
0.\ee Integrating this equation once, we have \be \dot u^2 + R^2
u^4 = C^2,\ee where $C^2$ is an integration constant which is
obvious non-negative and is actually proportional to the excess
energy (with a positive proportional constant) above the BPS
energy of the N supergravitons put on top of each other. It is
also obvious that $C^2 = 0$ gives the trivial $u = 0$ solution.
For $C > 0$, we have \be \label{u1} u(t) = - \sqrt{\frac{C}{R}} \mbox{cn}
[\sqrt{2 C R}(t - t_0)],\ee where $\mbox{cn} (x)$ is  one of Jacobian
elliptic functions. Like $\cos x$, it is a periodic function.  It has a real
period $4 K$ with $K$ determined by the modulus $k$ (here  $k^2 = 1/2$) as
\be\label{kp}
K = \int_{0}^{1} \frac{{\rm d}\, t}{\sqrt{\left(1 - t^{2}\right)\left(1 - k^{2} t^{2} \right)}}.
\ee
We have $-1 \le \mbox{cn} (x) \le 1$ and $\mbox{cn} (x + 2 K) = - \mbox{cn} (x), \mbox{cn} (x + 4 K) = \mbox{cn} (x)$, $\mbox{cn} (0) = 1, \mbox{cn} (K) = 0, \mbox{cn} (2 K) = - 1, \mbox{cn} (3 K) = 0$ and $\mbox{cn} (4 K) = 1$, some of which will be used later on.  For later use, we here also introduce the other Jacobian elliptic function  $\mbox{sn} (x)$ with $- 1 \le \mbox{sn} (x) \le 1$. It has the same real period as the $\mbox{cn} (x)$ given above but it is instead  like $\sin x$ with $\mbox{sn} (x + 2 K) = - \mbox{sn} (x), \mbox{sn} (x + 4 K) = \mbox{sn} (x), \mbox{sn} (0) = 0, \mbox{sn} (K) = 1, \mbox{sn} (2 K) = 0, \mbox{sn} (3 K) = - 1$ and $\mbox{sn} (4 K) = 0$.   Hence the above $u(t)$ given in (\ref{u1}) will oscillate between
values $- \sqrt{C/R} \le u(t) \le \sqrt{C/R}$. Therefore, we
expect a non-vanishing size if $C \neq 0$ (for example, the time
average of $u^2 (t)$ over a given period is non zero if $C \neq 0$
which is consistent with the spacetime uncertainty principle). To
make this a bit precise, let us follow \cite{myersone} to estimate
the growing size for a graviton as $\Delta L \sim   \sqrt{  < {\rm Tr}\,
(\Phi^i)^2>/N} \sim  N$ for large $N$ where we assume to take time
average, denoted as $< >$, in estimating the size. The total
excess energy can be estimated from the corresponding Hamiltonian obtained from the Lagrangian (\ref{bsp}) and gives $\Delta E \sim N^3 $. From this,
the excess energy for each degree of freedom (since we have U(N)
here) is therefore $\Delta E/N^2 \sim  N $. So we have the expected relation  $\Delta E \sim
\Delta L$ for $N \gg 1$ (we have chosen $l_p = 1$ conventions here).
We are certain that an excess energy without the presence of a
background flux can give rise to a finite size but there are still
some subtleties involved in having the precise uncertainty
relation. For examples, we are unable to explain why the time
average should be taken here and why the excess energy per degree
of freedom should be used.

\section{The dynamical spherical M2 brane}

    In this section, we want to find dynamical spherical M2 brane
    configurations from the DLCQ matrix theory of N
    super-gravitons in the pp-wave background which is obtained
     either from $AdS_4 \times S^7$ or from $AdS_7 \times S^4$ by taking the usual pp-wave limit.
    The action is given in \cite{bmnone} and we need only the
    bosonic part of the corresponding Lagrangian. The pp-wave background is
    \bea {\rm d}s^2 &=& - 2 {\rm d}x^- {\rm d} x^+ - \left[\left(\frac{\mu}{3}\right)^2(x_1^2 + x_2^2 + x_3^2)
    + \left(\frac{\mu}{6}\right)^2 (x_4^2 +  \cdots + x_9^2)\right] ({\rm d} x^{+})^{2} + {\rm d} {\vec x}^2\nn
     F_{+123} &=& \mu, \eea where we do DLCQ along the direction $x^- \sim x^- + 2\pi R$
     and we consider the sector of the theory with momentum $p^+ = - p_- = N/R$. With this,
     the bosonic part of the Lagrangian is
    \bea \label{lag} L &= &\frac{1}{2 R}\sum_{i = 1}^9 {\rm Tr}\,(\dot \Phi^i)^2 + \frac{R}{4} {\rm Tr}
    \sum_{i, j = 1}^9 [\Phi^i, \Phi^j]^2 + \frac{1}{2 R}{\rm Tr}\left[ -
    \left(\frac{\mu}{3}\right)^2\sum_{i = 1}^3 (\Phi^i)^2 -
    \left(\frac{\mu}{6}\right)^2 \sum_{j = 4}^9 (\Phi^j)^2\right]\nn
    &\,& -
    \frac{\mu}{3}\, {\rm i} \sum_{i, j, k = 1}^3 {\rm Tr}\, \Phi^i \Phi^j \Phi^k
    \epsilon_{ijk},\eea
    where we have taken $l_p = 1$, the gauge choice $A_0 = 0$
    and $t = x^+$.

   The equation of motion from the above Lagrangian is \bea\label{dc-eom}
   &&\ddot \Phi^i + R^2 \,[[\Phi^i, \Phi^j],\Phi^j] +
   \left(\frac{\mu}{3}\right)^2 \,\Phi^i + {\rm i}\, \mu R \,\Phi^j \,\Phi^k \,\epsilon_{ijk}
   = 0,\,\,(\mbox{for $i, j , k = 1, 2, 3$}),\nn
   && \ddot \Phi^l + R^2 [[\Phi^l, \Phi^m],\Phi^m] +
   \left(\frac{\mu}{6}\right)^2 \Phi^l = 0,\,(\mbox{for $l, m =
   4,\cdots,9$}),
   \eea
   where for the present interest and for simplicity we have
   assumed $[\Phi^i, \Phi^l] = 0$ and the repeated indices imply a summation.

   Let us consider the solutions of the second equation above,
   i.e., along  $l, m = 4, \cdots, 9$ directions. The trivial
   solution is obtained by setting $\Phi^l = g (t) \umatrix_{N\times N}$ with
   $\umatrix_{N\times N}$ the $N \times N$ unit matrix. We have $g (t) =
   g_0 \cos\mu (t - t_{0})/6$ with $g_0$ a constant, i.e., an oscillating
   solution because of the presence of the mass term. We have also
   non-trivial dynamical solutions, for example, similar to the one discussed
   in the previous solution, by choosing three of
   $\Phi^l$'s in the $N \times N$ irreducible representation of
   SU(2) as $\Phi^{3 + a} = g (t) K^a$ with $K^a$ the SU(2) generators for $a = 1, 2, 3$
   and the rest are trivial. Then we have the $g(t)$ satisfying
   the equation \be \ddot g + 2 R^2 g^3 + \left(\frac{\mu}{6}\right)^2 g =
   0,\ee which can be integrated once as \be \dot g^2 + R^2 \, g^4
   + \left(\frac{\mu}{6}\right)^2 g^2 = C^2,\ee with the integration constant
   $C^2 \ge 0$ and proportional to the contribution of the motion along these directions to
   the excess energy of the system under consideration. The
   solution is now \be \label{g} g (t) = - g_0 \mbox{cn} [\omega_0 (t -
   t_0)],\ee where we have \be g_0 =
   \sqrt{\frac{\sqrt{(\frac{\mu}{6})^4 + 4 R^2 C^2} -
   (\frac{\mu}{6})^2}{2 R^2}},\,\,\,\, \omega_0 =
   \left[\left(\frac{\mu}{6}\right)^4 + 4 R^2 C^2\right]^{1/4}.\ee
   This solution has the same characteristic as the one discussed
   in the previous section even though there is an additional mass term in
   the equation. It oscillates periodically with the modulus $k^2
   = R^2 g_0^2/\omega_0^2$ and between the value $g_0$ and $-
   g_0$. The solution collapses to a single point $g (t) = 0$ once
   $C^2 = 0$. Note also that this solution will reduce to the previous one when $R \to 0$ (so $k \to 0$) for which $\mbox{cn} (x) = \cos x$ and to the one given in the previous section when $\mu \to 0$.

   Now we come to discuss the first equation above along
    $i = 1, 2, 3$. Again we have a trivial solution
   $\Phi^i = u (t) \umatrix_{N\times N}$ with $u (t) = u_0 \cos \mu (t - t_{0})/3$. The non-trivial one can be obtained by setting
   $\Phi^i = u(t) J^i$ with $J^i$, the SU(2) generators, in the
   $N \times N$ irreducible representation of this group. Then we reduce the
   equation to the following \be\label{eom} \ddot u + 2 R^2 u^3 +
   \left(\frac{\mu}{3}\right)^2 u -  R \mu u^2 = 0.\ee Integrating this
   equation once, we have \be \dot u^2 + R^2 u^4 - \frac{2 R
   \mu}{3} u^3 + \left(\frac{\mu}{3}\right)^2 u^2= C^2,\ee again with the integration
    constant $C^2 > 0$ and proportional to the excess energy. $C^2 > 0$ can be easily
    seen if we re-express the above equation as \be \label{S-eom} \dot u^2 +
    R^2 u^2 \left(u - \frac{\mu}{3 R}\right)^2 = C^2.\ee For having  a clear physical picture
    and being convenient, we here give the relevant Lagrangian and the corresponding
   Hamiltonian. Since we have now $[\Phi^i, \Phi^l] = 0$, the $\Phi^i$ and
$\Phi^l$ with $i = 1, 2, 3$ and
    $\Phi^l = 4, \cdots 9$ are decoupled. Therefore, for the present purpose, we can simply focus on
    the Lagrangian, Hamiltonian etc for $\Phi^i$ only.  We denote
    the Lagrangian and Hamiltonian  as $\tilde L, \tilde H$
    respectively.  With our ansatz $\Phi^i = u(t) J^i$ with $J^i$ the
    generators of SU(2) in its irreducible $N \times N$
    representation, we then have \bea \label{L} \tilde L &=&\frac{1}{2 R} {\rm Tr}\, (\dot \Phi^i)^2 + \frac{R}{4} {\rm Tr}\,
    [\Phi^i, \Phi^j]^2 - \frac{1}{2 R}
    \left(\frac{\mu}{3}\right)^2 {\rm Tr}\,(\Phi^i)^2 -
    \frac{\mu}{3} \,{\rm i}\, {\rm Tr}\, \Phi^i \Phi^j \Phi^k \epsilon_{ijk} \nn
     &=& \frac{N (N^2 - 1)}{4} \left[\frac{1}{2 R} \dot u^2 - \frac{R}{2} u^4 + \frac{\mu}{3} u^3
     - \frac{1}{2 R} \left(\frac{\mu}{3}\right)^2 u^2\right],\eea
     where repeated indices mean summation and we have used ${\rm Tr}
     (J^i)^2 = N (N^2 - 1)/4$. The Hamiltonian is \be\label{H} \tilde H = \frac{N (N^2 -
     1)}{8 R} \left[ \dot u^2 + R^{2} u^2 \left(u -
     \frac{\mu}{3 R} \right)^2\right].\ee  For convenience, we define
     the reduced Hamiltonian $ H$ as \be \label{R-H} H \equiv
     \frac{8 R \tilde H}{N (N^2 - 1)} =  \dot u^2 + R^{2} u^2 \left(u -
     \frac{\mu}{3 R} \right)^2.\ee and the corresponding reduced
     potential as \be \label{p1} V = R^{2} u^2 \left(u -
     \frac{\mu}{3 R} \right)^2 \ge 0,\ee
     whose profile is shown in figure 1.

     \begin{figure}
     \begin{center}
     \includegraphics[scale=0.3]{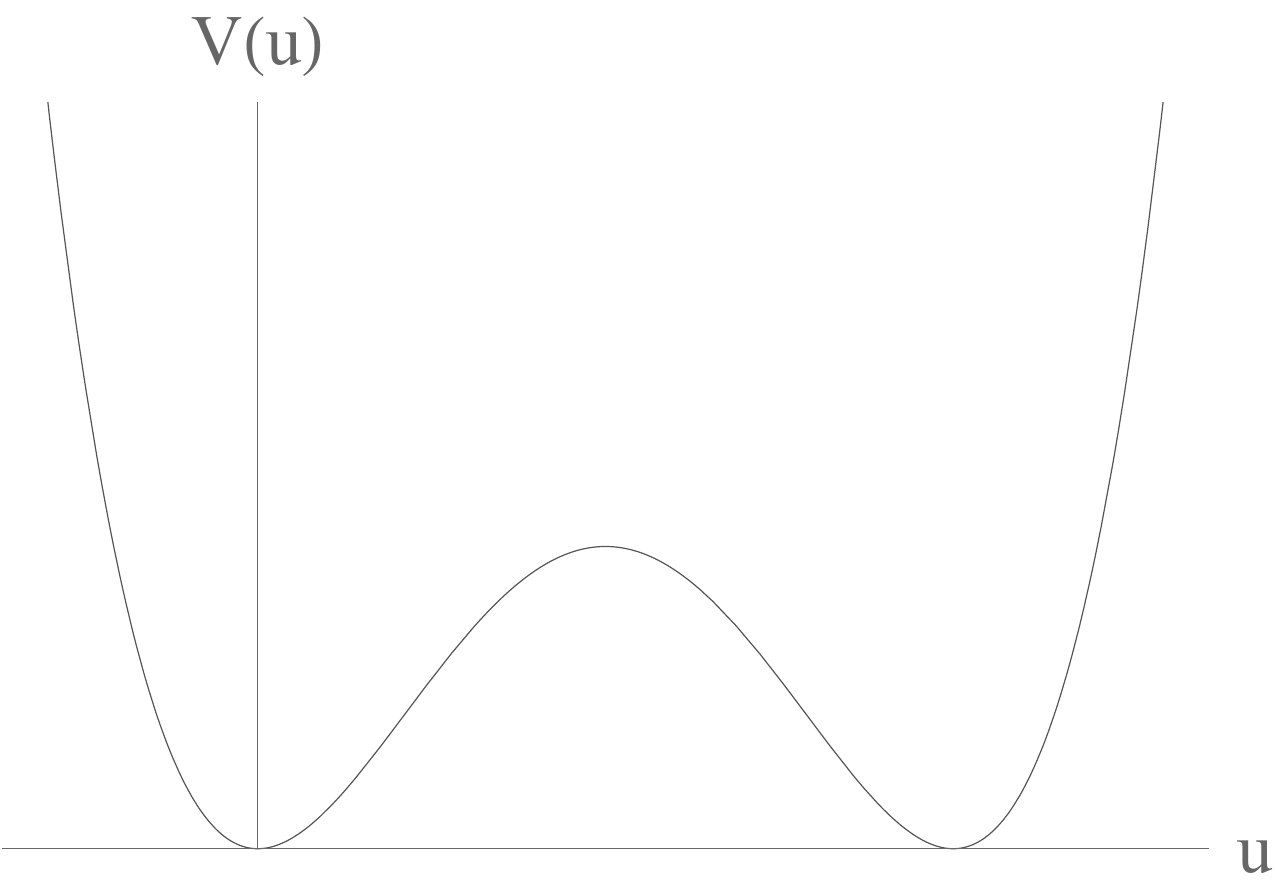}
     \end{center}
     \caption{\label{figone}The characteristic behavior of the potential (\ref{p1}). }
     \end{figure}

    Comparing (\ref{H}) and (\ref{R-H}) with (\ref{S-eom}), we have $\tilde H= N (N^2 - 1)
    C^2 /(8 R)$ and $ H = C^2$, both of which are  constant of
    motion as expected. Since $ V \ge 0$, therefore $u = 0$
    and $u = \mu/(3 R)$ give the minima of the potential which are both zero and
    therefore there must exist a $u$ between the two, which gives a local maximum of
    $V (u)$. It is not difficult to find this to be  $u = \mu/(6
    R)$ from ${\rm d} V/{\rm d} u = 0$. The corresponding value of the potential is
    $ V (u = \mu/(6 R)) = R^{2} (\mu/6 R)^4 $. These three points $u =0, \mu/(6 R),
    \mu/(3 R)$ satisfy the
    equation of motion (15) with $\dot u = 0$, therefore,
    the corresponding $C^2 = 0,  R^2 (\mu/6 R)^4, 0$, respectively.
     Given these, we know the profile
    of the potential, therefore the characteristic behavior of the
    solution from (\ref{S-eom}) for different $C^2$.

    The equation (\ref{S-eom}) cannot be solved in general using elementary functions and
    we have to use Jacobian elliptic function $\mbox{cn} (x)$ or $\mbox{sn} (x)$.
    For this purpose,
    we need to find four roots of the following polynomial
    \be P(u) = C^2 - R^2 u^2 \left(u - \frac{\mu}{3 R}\right)^2.\ee
    Depending on the range of the $C^2$ value, the corresponding solution will be
    different. We now discuss each such solution in order.

    {\bf Case (1)}: If $C = 0$, we can only have two static solutions $u = 0, \,
    \mu/3R$. $u = 0$ corresponds to the SUSY preserving BPS configuration of N
supergravitons while
    $u = \mu/3 R$ corresponds to a SUSY preserving BPS spherical M2 brane which is actually
    a giant as discovered in \cite{bmnone}. These two SUSY preserving solutions
    are expected from the profile of the potential and from our experience on giants.
     This case contrasts with what we discussed in
    \cite{clone} in which a dynamical spherical D2 exists which
    represents the creation of a semi-spherical D2 brane --anti
    semi-spherical D2 brane pair from N D0 branes and then
    annihilating back to N D0 branes at the end. The only SUSY
    preserving objects are the N D0 branes there.

    How to understand the presence of a SUSY preserving BPS giant here? First we have a pp-wave background which is
    obtained either from  $AdS_4\times S^7$ or from $AdS_7\times S^4$ and is
    different from the background employed in \cite{clone}. For $AdS_7 \times S^4$,
    there
    is, in addition to a SUSY preserving BPS super-graviton, a SUSY preserving BPS
    giant which is actually a spherical M2
    brane wrapped on a 2-sphere and orbiting on another $S^1$ in the $S^4$
    part with a given angular momentum. As noticed in \cite{gmtone}, both the
    zero-size graviton and the center of mass motion for the giant follow a null
    trajectory. For concreteness, let us write down  the line
    element for the $AdS_7 \times S^4$ as \be {\rm d} s^2 = L^2 (- {\rm d}t^2
    \cosh ^2 \rho + {\rm d}\rho^2 + \sinh^2\rho\, {\rm d}\Omega_5^2) + \frac{L^{2}}{4} ({\rm d}\phi^2
    \cos^2\theta + {\rm d}\theta^2 + \sin^2\theta \, {\rm d}\Omega_2^2),\ee
    where $L$ denotes the radius of $AdS_7$ and the radius of 4-sphere
    is $L/2$. Both the zero size graviton and the center of mass
    motion for the giant sit at $\rho = 0$ and $\theta = 0$. We
    also know \cite{bmnone} that the pp-wave geometry is the one
    near a null trajectory of a particle moving along the $\phi$
    direction with $\rho = 0$ and $\theta = 0$. Therefore, the
    geometry seen by either the zero-size graviton or the center of mass
    motion of the giant is the pp-wave. In obtaining a well-defined
    geometry, we need to send the $L \to \infty$. Those zero-size gravitons and finite
    size giants surviving under this limit correspond to the BPS
    configurations discussed here and already found in
    \cite{bmnone}. We here therefore provide an explanation about their
    existence. While for the background considered in
    \cite{clone}, we don't have a SUSY preserving BPS giant which
    can produce this background in a proper limit. Therefore, we
    don't expect a finite-size BPS configuration with zero excess
    energy there.

    The giant in the present case corresponds in certain sense to the dynamical
    spherical D2 brane with its size never vanishing in the case
    with a negative excess energy discussed in \cite{clone}. Unlike the present case,
    the D2 brane there is the only  allowed configuration.
    Classically, the present giant size can never be zero, analogous to the
    case for the D2 brane in \cite{clone}.  Quantum mechanically  here the story is different. The giant is
    itself a SUSY preserving BPS configuration and is stable. In
    appearance, it cannot turn itself to zero-size
    super-gravitons. However, it is well-known that this giant has
     all its properties, except for its non-zero size, identical to the
     zero-size supergraviton. In other words, we have two degenerate vacua
     in the present case, therefore an instanton tunnelling between the two is expected. So a finite size giant can turn itself into a zero size
    one  and vice-versa by this process. The instanton can actually be found in
    the usual way by sending $t \to - {\rm i} \tau$ with $\tau$ real. The corresponding Lagrangian $\tilde L_{\rm E}$ can be obtained from the Lagrangian in (\ref{L}) by the standard procedure as 
    \bea \tilde L \to \tilde L_{\rm E} &=& - \tilde L (t \to - {\rm i} \tau)\nn
    &=& \frac{N (N^{2} - 1)}{8 R} \left[ u'^{2} + R^{2} u^{2} \left(u - \frac{\mu}{3 R}\right)^{2}\right],
    \eea
    where we denote $u' \equiv {\rm d} u/{\rm d}\tau$. From this, we have the corresponding Hamiltonian $\tilde H_{\rm E}$ as
    \be
    \tilde H \to \tilde H_{\rm E} = - \tilde H(t \to - {\rm i}\tau) = \frac{N(N^{2} - 1)}{8 R} \left[u'^{2} -  R^{2} u^{2} \left(u - \frac{\mu}{3 R}\right)^{2}\right].
    \ee
    The corresponding reduced Hamiltonian 
    \be
    H \to H_{\rm E} = - H (t \to - {\rm i} \tau) = u'^{2} -  R^{2} u^{2} \left(u - \frac{\mu}{3 R}\right)^{2},
    \ee
    which gives the reduced potential now as
    \be\label{rdp}
    V_{\rm E} (u) = - R^{2} u^{2} \left(u - \frac{\mu}{3 R}\right)^{2},
    \ee
    which differs from the real time one by a sign (see figure 2 for its profile).   
    
     \begin{figure}
     \begin{center}
     \includegraphics[scale=0.4]{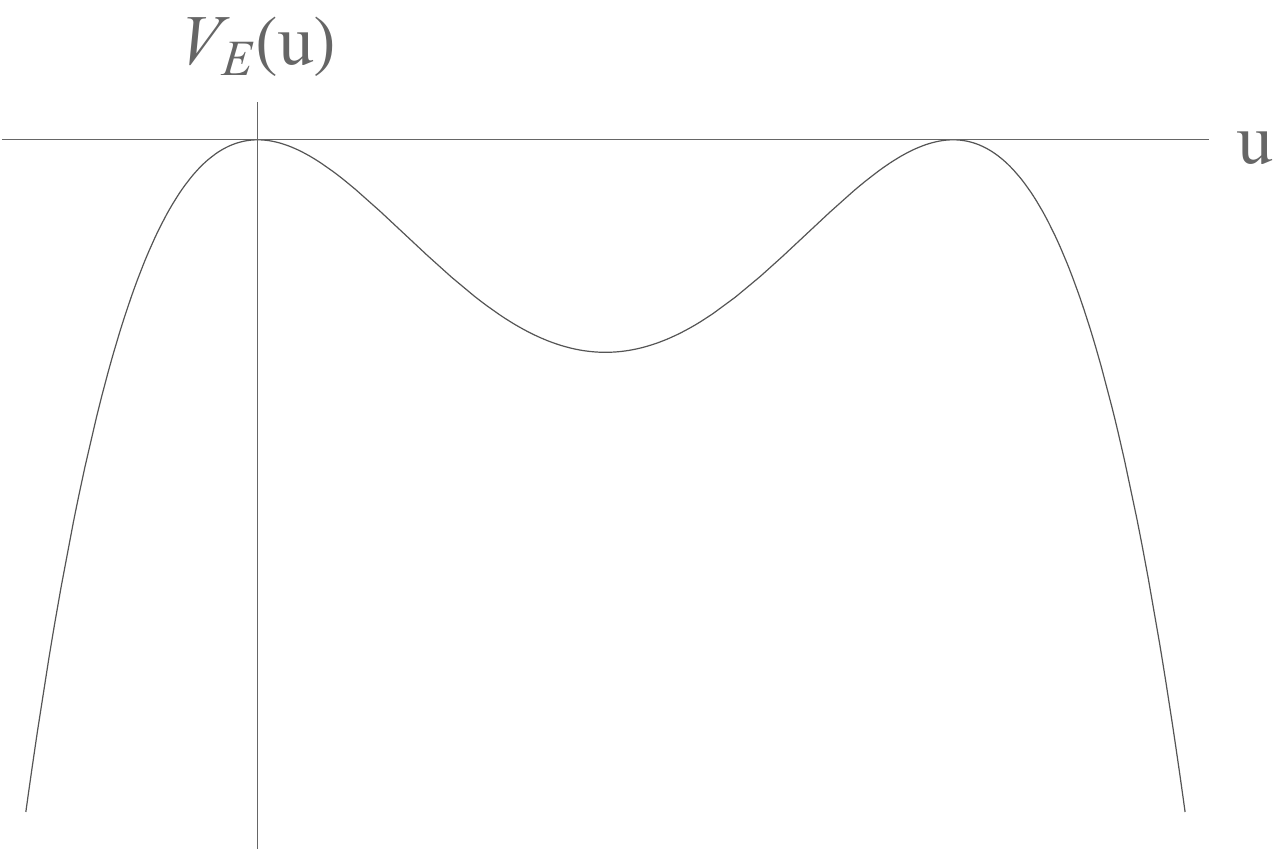}
     \end{center}
     \caption{\label{figtwo}The characteristic behavior of the reduced potential (\ref{rdp}). }
     \end{figure}
    
     In other words, the instanton has its dynamics in imaginary time.
    The equation of motion satisfying by the instanton is \be u'^2
    - R^2 u^2 \left(u - \frac{\mu}{3 R}\right)^2 = - C^2, \ee with the same
    $C^2$, which can also be obtained from eqn. (\ref{S-eom}) by sending $t\to - {\rm i} \tau$. We expect an instanton tunnelling between the BPS graviton and the BPS giant for both of which $C = 0$.   For the case $C = 0$, the above
    equation can be solved easily as \be u(\tau) = \frac{\mu}{3 R}
    \frac{1}{1 + {\rm e}^{- \mu(\tau - \tau_0)/3}},\ee where we have chosen
    $u (\tau = - \infty) = 0$ and $u(\tau = \infty) = \mu/(3 R)$.
    When we choose $u (\tau = - \infty) = \mu/(3 R)$ and $u(\tau = \infty) =
    0$, the instanton profile is \be u(\tau) = \frac{\mu}{3 R}
    \frac{1}{1 + {\rm e}^{\mu(\tau - \tau_0)/3}}.\ee  In either of the two cases above, we also choose $u = \mu/6R$ when $\tau = \tau_{0}$ which is  reasonable\footnote{One can always do so by choosing a proper $\tau_{0}$.} since the potential is symmetric with respect to this particular $u = \mu/6 R$. The WKB tunnelling
    amplitude can be evaluated as $\sim e^{- \Delta S_E}$ with the action change
    $\Delta S_E$ as
    \bea
    \Delta S_E &=& \int_{- \infty}^\infty \left( \tilde L_E (\tau) - \tilde L_E (\mbox{end points})\right)
    {\rm d}\tau,\nn
    &=& \frac{N(N^2 - 1)}{3}\left(\frac{\mu}{6 R}\right)^3,\eea
    for the instanton profile, noting that $\tilde L_{\rm E} (\mbox{end points}) = 0$.  So for large $N$, the transition
    probability is vanishingly small.  This is consistent with the fact that the size of the giant is larger and the potential barrier at $u = \mu/(6 R)$ is 
    \be
    \tilde V (u = \mu/(6 R)) = \frac{\mu N (N^{2} - 1)}{48} \left(\frac{\mu}{6 R}\right)^{3},
    \ee
    is higher for larger $N$.

    {\bf Case (2):} If $0 < C < R(\mu/ 6 R)^2$, we have four real
    roots for $P(u) = 0$ which is obvious from figure 1. We have two disconnected dynamical
    configurations, either of which resembles the periodical motion in the case of
    no background flux as discussed in section 2 but now the
    size in one case can be zero while the size in the
    other can never be zero. The present case has some similarity, for example, for the non-zero size case, to the negative excess energy case
    discussed in \cite{clone} but also has many differences, for example, we have an
    additional branch of the solution whose size can be zero. This is entirely due to that we have
    two BPS configurations when $C = 0$, one has zero size while the other has finite size, the giant.
    The four real
    roots can be found to be \bea a &=& \frac{\mu + (36 R C + \mu^2)^{1/2}}{6
    R},\,\,\, b = \frac{\mu + (- 36 R C + \mu^2)^{1/2}}{6
    R},\nn
    c &=& \frac{\mu - (- 36 R C + \mu^2)^{1/2}}{6 R},\,\,\,
    d = \frac{\mu - (36 R C + \mu^2)^{1/2}}{6 R},\eea where
    $a > b > c > d$ and with $d < 0$.

    With the above four roots, the dynamical oscillating configuration
    containing  $u = 0$, i.e.,  for
    which $d \le u (t) \le c$, is given as \be\label{2-1sol} u(t) = \frac{(a - c) d + (c - d)
    a\, \mbox{sn}^2 ( R (t - t_0)/g)}{(a - c) + (c - d)\, \mbox{sn}^2
    (R (t - t_0)/g)},\ee where $\mbox {sn} (x)$ is a Jacobian
    elliptic function which, like $\sin (x)$, is a periodic
    function and has a range of $-1 \le \mbox{sn} (x) \le 1$ with
    its period determined by the modulus $k^2$. Both of the $k^2$ and the parameter
    $g$ appearing in the above are given in terms of the four roots as
    \be \label{gk} g = \frac{2}{\sqrt{(a - c)(b - d)}},\,\,\,\, k^2 =
    \frac{(a - b)(c - d)}{(a - c) (b - d)}.\ee It is not difficult
    to check that as $\mbox{sn}^2 (R (t - t_0)/g) = 0$, we have $u
    (t) = d$ while $\mbox{sn}^2 (R (t - t_0)/g) = 1$, $u (t) = c$
    as expected.

     The other dynamical oscillating configuration containing the value
      $u = \mu/(3 R)$, i.e., for which $ b \le u \le a$, is
     given as \be\label{2-2sol} u (t) = \frac{(b - d)a + (a - b) d\, \mbox{sn}^2
     (R (t - t_0)/g)}{(b - d) + (a - b)\, \mbox{sn}^2 (R (t -
     t_0)/g)},\ee where both the parameter $g$ and the modulus
     $k^2$ used to determine the period are the same as the ones
     given above. When $\mbox{sn}^2 (R(t - t_0)/g) = 0$, we
     have $u(t) = a$ while $\mbox{sn}^2 (R(t - t_0)/g) = 1$, $u(t)
     = b$, again as expected.

     In either of the above two cases, the period of $u (t)$ is only half of that of $\mbox{sn} \left(R (t - t_{0})/g\right)$, i.e., $2 K$, with $K$ determined by (\ref{kp}) but now with
     the modulus given in (\ref{gk}).
     
     Now considering the limiting case $C \to
     0$, we have now $a \to b \to \mu/3 R$ and $c \to d \to 0$. We
     therefore have $g \to 6 R/\mu$ and $k^2 \to 0$. From the
     solution of $d \le u (t) \le c$, we have $u (t) = 0$ and from the
     solution of $b \le u (t) \le a$ we have $u(t) = a = \mu/3 R$,
     i.e., either collapses to the corresponding SUSY preserving BPS
     configuration as expected.

     {\bf Case (3):} For $C = R (\mu/6R)^2$, this serves as a
     transition place between the solutions discussed in Case (2)
     and the one which will be discussed after the present case.
     This is also analogous to the zero-excess-energy case discussed in \cite{clone}
     for which the
     solution is expected to be representable by elementary
     functions but as we will see that the behavior of the 
     present solutions is different from the one discussed in
     \cite{clone}. We also have two solutions here instead of one.
     This is also due to the fact that we have
     two BPS configurations, one with zero size while the other
     with finite size, when $C = 0$. We now have three different real roots
     since now the roots $b$ and $c$ given in (26) of Case (2) become
     degenerate. They are now \be a = \frac{(1 + \sqrt{2})\mu}{6
     R},\,\,\, b = c = \frac{\mu}{6 R},\,\,\, d = \frac{(1 -
     \sqrt{2})\mu}{6 R},\ee where the root $d < 0$ while the rest
     are positive. We have for $ d \le u(t) \le c$, \be\label{P} u(t) =
     \frac{\mu}{6 R} \left[1 - \frac{\sqrt{2}}{\cosh[\frac{\mu(t -
     t_0)}{3\sqrt{2}}]}\right],\ee which represents that at $t = -
     \infty$, $u(t) = \mu/(6 R)$, then the size becomes smaller
     and reaches zero at a point $\cosh [\mu(t - t_0)/(3\sqrt{2})]
     = \sqrt{2}$, then $u (t) < 0$ and reaches  $u = -
     (\sqrt{2} - 1)\mu/(6 R) = d $ at $ t = t_0$. Further, it
     reaches zero-size again and finally returns to $u =
     \mu/(6 R) = c$ at $t = \infty$. For $b \le u \le a$, we
     have\be\label{N} u (t) = \frac{\mu}{6 R}\left[1 +
     \frac{\sqrt{2}}{\cosh[\frac{\mu(t -
     t_0)}{3\sqrt{2}}]}\right].\ee  At $t = \pm \infty$, we have
     $u = \mu/(6 R) = b$ while at $t = t_0$, $u = (1 + \sqrt{2})
     \mu/(6 R) = a$. Apart from the point $u = a$, every other
     size will be reached twice when the time evolves from $t = -
     \infty$ to $+\infty$.

     One can check that the two solutions can be obtained from the
     corresponding two discussed in Case 2
      by taking the limit $C \to R (\mu/6 R)^2$ there, noting that now $k \to 1$ and $\mbox{sn} (x) \to \tanh (x)$. 

     {\bf Case (4):} If $C > R (\mu/6 R)^2$, we have two real roots $a$
     and $b$ with $a > b$ and two complex roots $c$ and $\bar c$
     which are complex conjugate to each other. They are given as
     follows:\bea \label{roots-4} a &=& \frac{\mu + (36 R C + \mu^2)^{1/2}}{6
    R},\,\,\, b = \frac{\mu - (36 R C + \mu^2)^{1/2}}{6
    R},\nn\\
    c &=& \frac{\mu +{\rm i}\,( 36 R C - \mu^2)^{1/2}}{6 R},\,\,\,
    \bar c = \frac{\mu - {\rm i}\,(36 R C - \mu^2)^{1/2}}{6 R}.\eea
    Only for this case, we have a solution whose characteristic behavior
    is the same as the one obtained in \cite{clone} for positive excess energy  there.
    With the above four roots, we have the periodically oscillating
     solution as \be \label{sol-4} u(t) = \frac{(a B + b A) - (a B - b A)
     \mbox{cn} [R (t - t_0)/g]}{(A + B) + (A - B)\mbox{cn} [R(t -
     t_0) /g]},\ee where the parameters $A, B, g$ and the modulus
     $k^2$ are defined as $ A^2 = (a - b_1)^2 + a_1^2, B^2 = (b -
     b_1)^2 + a_1^2, g = 1/\sqrt{AB}, k^2 = [(a - b)^2 - (A -
     B)^2]/(4 A B)$ with $a_1^2 = - (c - \bar c)^2/4, b_1 = (c +
     \bar c)/2$. We can actually simplify the above solution (\ref{sol-4}) when the explicit roots (\ref{roots-4}) are used. We have
     now 
     \be
     a^{2}_{1} = \frac{36 RC - \mu^{2}}{36 R^{2}}, \quad b_{1} = \frac{\mu}{6 R}.
     \ee
     So we have
     \be
     a - b_{1} = \frac{\left(36 R C + \mu^{2}\right)^{1/2}}{6 R} = - (b - b_{1}),
     \ee
     which implies 
     \be
     A = B = \sqrt{\frac{2 C}{R}}.
     \ee  
     Now 
     \be\label{gk-4}
     g = \frac{1}{A} = \sqrt{\frac{R}{2 C}}, \quad k^{2} = \frac{1}{2} \left(1 + \frac{\mu^{2}}{36 RC}\right) < 1.
     \ee
     With this, we have the solution (\ref{sol-4}) as
     \be\label{4-sol}
     u(t) = \frac{(a + b)}{2} - \frac{(a - b)}{2} \mbox{cn} \left[\sqrt{2 RC} (t - t_{0})\right].
     \ee
     One can check that $u = b$ when $\mbox{cn} [R(t -
     t_0) /g] = 1$ (i.e., when $\sqrt{2 CR} (t - t_{0}) = 4 K n$ with $n = 0, \pm 1, \pm 2, \cdots$ and $K$ determined by (\ref{kp}) with the present modulus given in (\ref{gk-4}) ) while $u = a$ when $\mbox{cn} [\sqrt{2 CR} (t -
     t_0)] = -1 $ (i.e., when $\sqrt{2 CR} (t - t_{0}) = 2 K n$).  Note that when $C \to R (\mu/6 R)^{2}$, $k^{2} \to 1$ and $K \to \infty$ from (\ref{kp}). We expect now the present solution reduces to the two solutions given in the previous case.  Since now $K \to \infty$, we need to do this carefully in order to get the correct results. We only need to consider one period of $\mbox{cn} (x)$ and the proper choice is $- 2 K \le x \le 2 K$. For the branch of $\mbox{cn} (x) \ge 0$, it is easy to realize and we just limit $ - K \le x \le K$ but for the branch of $\mbox{cn} (x) \le 0$, this is for $- 2 K \le x \le - K$ and $ K \le x \le 2 K$. For the latter branch, extra effort is needed. For $- 2 K \le x \le - K$, we define $y = x + 2 K$ with $0 \le y \le K$  and have $\mbox{cn} (x) = \mbox{cn} (y - 2 K) = - \mbox{cn} (y)$. For $K \le x \le 2 K$, we define $y = x - 2 K$ with $ - K \le y \le 0$ and have $\mbox{cn} (x) = \mbox{cn} (y + 2 K) = - \mbox{cn} (y)$.  In other words, we can use $- K \le x \le K$ for both branches and  use $\mbox{cn} (x)$ for the original positive branch and $- \mbox{cn} (x)$ for the negative branch. Noting that $\mbox{cn} (x) \to 1/\cosh (x)$ when $k^{2} \to 1$ and $\sqrt{2 C R} = \mu/3\sqrt{2}$. It is easy to check that the positive branch gives (\ref{P}) while the negative branch gives (\ref{N}) in the previous case. 

\section{Discussions}

      We give a rather detail analysis of the dynamics for N supergravitons, which can be described by DLCQ matrix theory, moving in the pp-wave
       background
      obtained either from $AdS_4 \times S^7$ or from $AdS_7
      \times
      S^4$. We find differences from and similarities to what was
      discovered about the dynamics of N D0 branes moving in a
      flat background with a RR 4-form flux in\cite{clone}. In addition to what
      have been said in the text, the differences are mainly due
      to that between the two backgrounds, the former
      preserves the maximal SUSY while the later doesn't preserve
      any. The quantity which characterizes the differences as well as the
      similarities  is the
      parameter $C$ which is related to the so-called excess
      energy. The BPS configurations can be reached in both cases
      for $C = 0$. In the present case, we have two, one is the
      zero-size BPS graviton and the other the finite size giant,
      which again are due to the maximal SUSY preserving pp-wave
      background. While for the case discussed in \cite{clone}, the best we
      can have is the BPS D0 branes. The differences occur only
      for the excess energy below some critical value even though there are
      still similarities there as discussed in the text. Above
      that value, the characteristic behavior is essentially the
      same in both cases. This seems to give hints that vacua
      have important influences on the low energy dynamics but don't do
      that much on the high energy one whose characteristic
      behavior seems independent of the vacuum chosen.

      The other important question\footnote{We thank one of our  anonymous referees for asking us to give a discussion on this issue.} is about the stability issue for each of the time-dependent solutions found in this paper.   However, this is a very complicated issue whose complete answer is well beyond the scope of this paper. The instability here consists of classical and quantum ones.  For the classical one, we have linear and non-linear ones, while for quantum one, we have perturbative and non-perturbative ones.  For example,  both the BPS ground states considered in this paper, i.e., N super-gravitons and the giants, are obviously stable not only classically but also quantum-mechanically in perturbation.  
As shown earlier,  for finite N,  these two BPS vacua can turn into each other non-perturbatively  via instanton tunnelling, due to their degeneracy in energy.  So only for large N, either of them is indeed stable under all cases mentioned above. 

In the following, for the time-dependent solutions found in this paper, we only give a very limited discussion about the classical linear instability, due to the complexity of the issue.  If we limit ourselves to the assumptions made for the time-dependent solutions discussed in the paper also for the linear perturbation, i.e., $u (t) = u_{0} (t) + \delta u (t)$ where $u_{0} (t)$ is one of the time-dependent solutions given in section 3, we would expect the following.   For case (2),  if the excess energy is not close to the one  for case (3) (called it the critical one), even though we have two degenerate solutions (\ref{2-1sol}) and (\ref{2-2sol}), we don't expect that the two  
can turn into each other under small perturbation, due to that there is a high potential barrier between the two and the present system as given by the Lagrangian (\ref{lag}), unlike in \cite{clone},  can be viewed as an isolated one, therefore conserving the system energy.  For case (4), we don't even have degenerate solutions and expect that the time-dependent solution (\ref{4-sol}) is stable under a small linear perturbation if the excess energy is not close to the critical one mentioned above for the same reasons.  The only cases we need to worry about are the case (3),  and the case (2) and case (4) when their respective excess energy is close to the critical one.  For these cases, the time-dependent solutions can be (or be given approximately) as (\ref{P}) or as (\ref{N}).  We in the following choose either of them as $u_{0} (t)$.

   Under a small perturbation $\delta u (t)$, the linearized perturbation equation from (\ref{eom}) is as
     \be\label{small-perb}
     \delta \ddot u + \left[ 6 R^{2} \left(u_{0} (t) - \frac{\mu}{3 R}\right) u_{0} (t) + \left(\frac{\mu}{3}\right)^{2} \right] \delta u = 0,
     \ee
    which becomes 
    \be\label{small}
    \delta \ddot u - \frac{1}{2} \left(\frac{\mu}{3}\right)^{2} \left[1 - \frac{6}{\cosh^{2} \left[\frac{\mu (t - t_{0})}{3 \sqrt{2}}\right]}\right] \delta u = 0,
    \ee
    where we have used the solution (\ref{P}) or (\ref{N}).
    
    When $t \to \pm \infty$, $u_{0} (t) \to b = c = \mu/(6 R)$ and we expect that the instability can arise since this is the place where the two degenerate solutions meet.  When $t \to t_{0}$,  the solution (\ref{P}) is close to $u_{0} (t_{0}) = d = - (\sqrt{2} - 1) \mu/(6 R)$ while the solution (\ref{N}) is close to  $u_{0} (t _{0}) = a = (1 + \sqrt{2})\mu/(6 R)$. Since $a > 0$ and $d < 0$ are far apart, we don't expect that the two can turn into each other under a small perturbation. The perturbation equation (\ref{small}) reflects what we have just described.  Let us see the detail. When $t \to \pm \infty$, the equation (\ref{small}) becomes
  \be
  \delta \ddot u - \frac{1}{2} \left(\frac{\mu}{3}\right)^{2} \delta u = 0,
  \ee
  which has two solutions
  \be
  \delta u \sim {\rm e}^{\pm \frac{\mu t}{3 \sqrt{2}}},
  \ee
  for which one of the modes (with the plus sign) is exponentially growing, signalling the instability of the time-dependent solution $u_{0} (t)$ (\ref{P}) or (\ref{N}) at large $|t|$. While for small $t - t_{0}$, we have from (\ref{small})
  \be
  \delta \ddot u + \frac{5}{2} \left(\frac{\mu}{3}\right)^{2} \delta u = 0,
  \ee
  which gives a solution
  \be
  \delta u  = a_{0} \sin \left[\frac{\mu}{3} \sqrt{\frac{5}{2}} \left( t - t_{0}\right) +  \phi_{0}\right],
  \ee
  an oscillating mode which can remains small for small $a_{0}$, therefore indicating that the original solution $u_{0} (t)$ is stable at small $t - t_{0}$. Given that the solution (\ref{P}) or $(\ref{N})$ is not stable at large $|t|$, so overall either solution is unstable, as expected. 
  
  In other words, any of the time-dependent solutions given in section 3 is unstable even under a small perturbation when their respective excess energy is or is close to the critical one.
  The above analysis appears to  imply that the time-dependent solutions given in section 3 could be stable under small perturbations if the respective excess energy is away from the critical one. This is actually far from clear if a general perturbation is considered, i.e., not the one considered above. In general,  for any of the time-dependent solutions $u_{0} (t)$ given in section 3 and the solution $g(t)$ given by (\ref{g}), we have
  \be
  \Phi^{i} (t) = u_{0} (t) J^{i}  + \delta \Phi^{i}, \qquad  \Phi^{l} (t) =  g (t) \umatrix_{N\times N} + \delta \Phi^{l},
  \ee
  where $i = 1, 2 , 3$ and $l = 4, 5, \dots, 9$.  In general, both $\Phi^{i}$ and $\Phi^{l}$ are U(N) matrices and so do $\delta \Phi^{i}$ and $\delta \Phi^{l}$. In other words, both $\delta \Phi^{i}$ and $\delta \Phi^{l}$ can each give rise to $N^{2}$ modes.  In general, we can also have $[J^{i}, \delta \Phi^{l}] \neq 0$. So under linear perturbations, the corresponding equations are in general coupled ones, which should be derived from the original Lagrangian (\ref{lag}), though $u_{0} (t)$ and $g (t)$ are from the decoupled ones as given in (\ref{dc-eom}). As expected, the analysis of such coupled perturbations with general $\delta \Phi^{i}$ and $\delta \Phi^{l}$ are hard. Though the system under consideration is indeed isolated and the total energy is conserved, the excess energy associated with each of the time-dependent solutions given in section 3 and that with the solution $g (t)$ as given in (\ref{g}) can however dissipate to excite those modes which are so far frozen for the solutions considered in section 3.   That is to say that the time-dependent solutions found in this paper can be potentially unstable and as mentioned earlier a complete analysis of them is far  beyond the scope of this paper and is worth some other independent  effort. 
       
      We also provide evidence in support of spacetime uncertainty
      principle by showing that the size growth of particles or
      extended object is indeed related to the excess energy even
      without the presence of a background flux.
      
      In the present paper, we only consider the simplest cases for which we limit ourselves to the non-trivial N-dimensional representation of SU(2) in either 
      i-directions or $l$-directions. We could extend this to a more general case in both directions, for example, considering SU(2) $\times$ SU(2), with one SU(2) in i-directions and the other one in $l$-directions, but still insist $[\Phi^{i}, \Phi^{l}] = 0$ for simplicity.  This may give rise to some interesting cases which may be worth a try.

\vs{5}

\noindent {\bf Acknowledgements}

\vs{2}
We thank one anonymous referee for a question regarding the instability issue of the time-dependent solutions found in this paper and the other for pointing out one repeated reference to improve the manuscript.  We acknowledge support by grants from the NSF of China with Grant No: 11775212 and 12047502.


\end{document}